\documentclass[aps,pra,reprint,superscriptaddress, showpacs]{revtex4-1}

\usepackage{graphicx}
\usepackage{dcolumn}
\usepackage{bm}
\usepackage{xcolor}
\usepackage{longtable}
\usepackage{amsmath}

\begin{document}


\title{Atom interferometry using a shaken optical lattice}


\author{C.A. Weidner}
\affiliation{Department of Physics and JILA, University of Colorado, Boulder, Colorado, 80309-0440, USA}
\author{Hoon Yu}
\altaffiliation[Current address:~]{Hanwha Corporation Defence R\&D center, Daejeon, Republic of Korea}
\affiliation{Department of Physics and JILA, University of Colorado, Boulder, Colorado, 80309-0440, USA}
\author{Ronnie Kosloff}
\affiliation{Fritz Haber Research Centre and The Department of Physical Chemistry, Hebrew University, Jerusalem 91904, Israel}
\author{Dana Z. Anderson}
\affiliation{Department of Physics and JILA, University of Colorado, Boulder, Colorado, 80309-0440, USA}
\email{dana@jila.colorado.edu}


\date{\today}

\begin{abstract}
We introduce shaken lattice interferometry with atoms trapped in a one-dimensional optical lattice. By phase modulating (shaking) the lattice, we control the momentum state of the atoms. Through a sequence of shaking functions, the atoms undergo an interferometer sequence of splitting, propagation, reflection, reverse-propagation and recombination. Each shaking function in the sequence is optimized with a genetic algorithm to achieve the desired momentum state transitions. As with conventional atom interferometers, the sensitivity of the shaken lattice interferometer increases with interrogation time. The shaken lattice interferometer may also be optimized to sense signals of interest while rejecting others, such as the measurement of an AC inertial signal in the presence of an unwanted DC signal.
\end{abstract}

\pacs{37.25.+k,37.10.Jk, 03.75.Dg}

\maketitle


\section{\label{Sec.Intro} Introduction}

This work introduces an approach to atom interferometry using a shaken optical lattice. Consider atoms interacting with an off-resonant one-dimensional standing light field produced by retro reflecting an incoming laser beam with a mirror. By ``shaken'' we mean that the longitudinal positions of the lattice nodes are modulated, for example, by moving the reflecting mirror back and forth. The pioneering work of P{\"o}tting et al. \cite{Meystre2001} established that it is possible to transform an initial atomic wave function into another desired wave function by shaking the lattice in a prescribed way. The shaken lattice concept has also been broadly applied to study atom tunneling and transport in lattices \cite{Ferrari2009, Arimondo2007, Arimondo2009} as well as ferromagnetism \cite{Chin2013}. Atoms held in a shaken lattice have also been used to measure gravity \cite{Tino2008, Tino2012}.

\begin{figure}[]
\includegraphics[scale = 0.335]{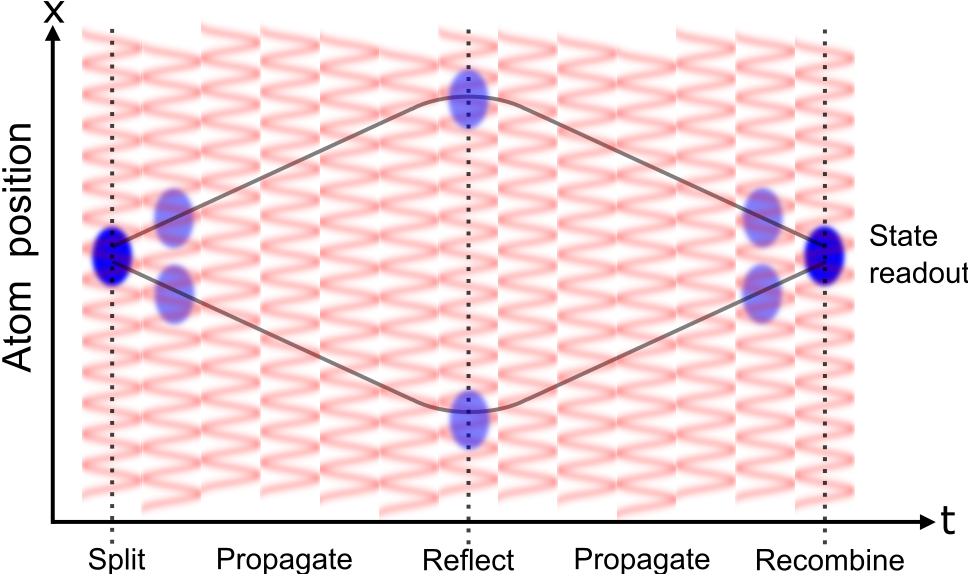}
\caption{\label{inter_full} (color online) The full interferometer sequence. Blue clouds represent atom wavepackets interacting with the shaken lattice. Atoms begin in the ground Bloch state of the lattice (Fig. \ref{fig.init}) and are split into two oppositely propagating wavepackets. The atoms are then reflected, reverse-propagated, and recombined back into their initial ground state (in the absence of a signal), thus completing the interferometer sequence.}
\end{figure}

In their original work, P{\"o}tting et al. sought a specific redistribution of the momentum state of atoms in the lattice. They achieved the desired wavefunction transformation through the use of a genetic algorithm (GA) \cite{Meystre2001, Rabitz1992, Haupt2004}. Here we use this technique to carry out a set of transformations that reproduce the ordered sequence of operations associated with a Michelson interferometer \cite{Wu2005}, namely splitting, propagation, reflection, reverse-propagation, and recombination of the atomic wavefunction, as shown in Fig. \ref{inter_full}. The protocol needed to execute each operation is found through the use of a GA. While we use a GA to optimize the shaking, the interferometer sequence may also be found through use of an optimal control algorithm such as the Krotov \cite{Kosloff2003, Kosloff2008} or CRAB methods \cite{Montangero2011}. Similar methods have been used to solve problems in other atomic systems, such as matter wave pulse shaping \cite{Kosloff2009} or state inversion of a BEC \cite{Hohenester2013}. Optimization methods have been used in atom interferometers, as in the Ramsey interferometry scheme of reference \cite{Schmiedmayer2014} or to find efficient light pulse schemes for atom interferometry \cite{Close2012}.

\begin{figure}[]
\includegraphics[scale = 0.265]{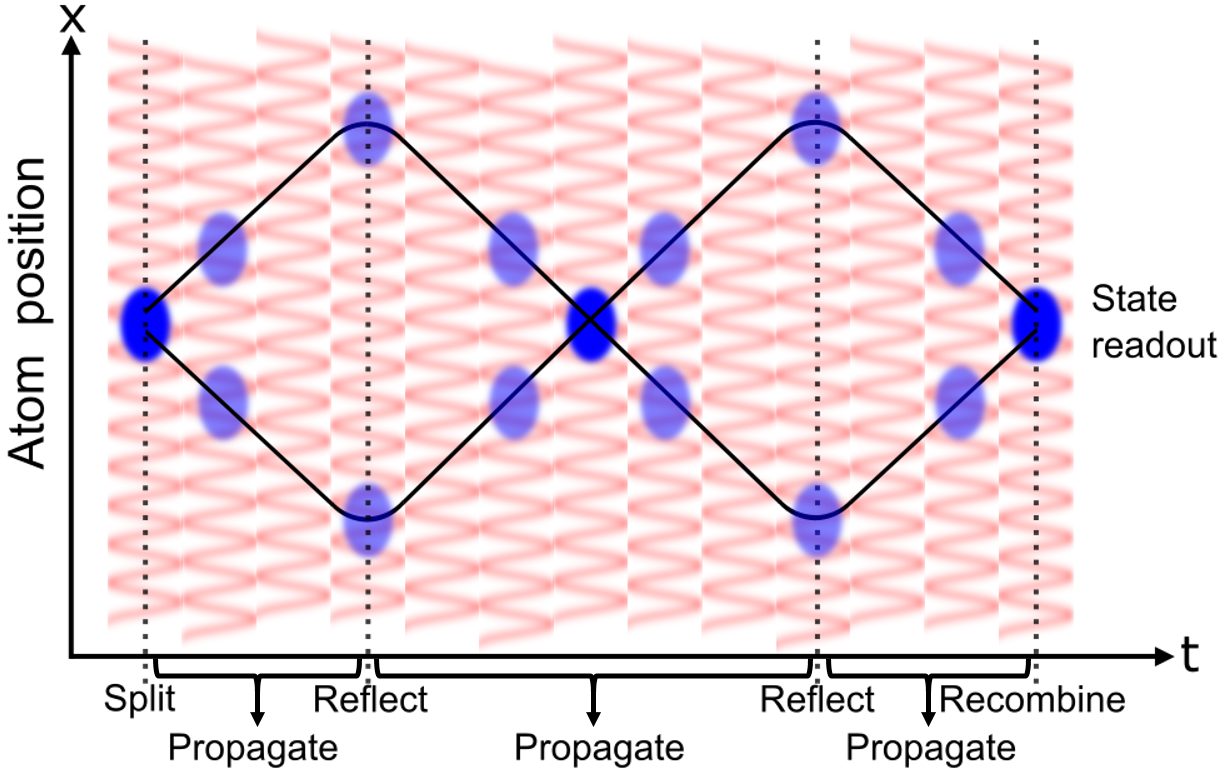} 
\caption{\label{fig.inter_recip} (color online) The reciprocal interferometer sequence. The reciprocal interferometer modifies the standard Michelson interferometer sequence shown in Fig. \ref{inter_full} so that the atoms travel a fully symmetric path. This configuration is designed to be sensitive to AC accelerations and immune to DC accelerations.}
\end{figure}

We show that in the Michelson configuration, the interferometer's sensitivity to acceleration is consistent with the $T_\mathrm{I}^2$ sensitivity achieved by free-space atom interferometers \cite{Chu1991, Kasevich1997, Kasevich2000, Bouyer2006, Kasevich2013}. Here, $T_\mathrm{I}$ is the interrogation time of the interferometer. Furthermore, one interesting aspect of shaken lattice interferometry (SLI) that is distinguishable from conventional atom interferometry is that SLI affords a means to control the transfer function of the system. As a simple example consider a scenario in which one wishes to eliminate the interferometer sensitivity to the constant force of gravity. By optimizing the interferometry sequence in the presence of gravity, this DC signal may be eliminated. Additionally, the interferometer may be configured to be sensitive to AC signals. Using the reciprocal scheme shown in Fig. \ref{fig.inter_recip}, a DC signal may be eliminated in favor of sensitivity to an AC signal. Thus, using SLI one can tailor the sensitivity of the interferometer to a given signal by learning to enhance a desired signal and reject undesired signals outside of the target sensitivity range.

This work proceeds as follows: In Sec. \ref{sec.Opt} the numerical simulation and optimization process is presented. Section \ref{sec.Mich} discusses how the optimization is applied to the standard Michelson interferometer. Section \ref{sec.Recip} shows how a reciprocal scheme may be used to enhance an AC acceleration signal and suppress a DC signal. Section \ref{sec.Sensitivity} gives a detailed analysis of the sensitivity of the shaken lattice interferometer as well as the errors that could affect measurements made in an experimental system. Section \ref{sec.Robustness} analyzes the robustness of the results in the presence of variations in parameters, and Sec. \ref{sec.Conc} concludes.

\section{\label{sec.Opt}Simulation and optimization of the shaken lattice interferometer}

In this section, we provide an overview of the SLI concept and present the numerical and optimization methods we use to study it. We assume that non-interacting atoms are trapped in an infinite 1-D lattice potential and study their evolution using the one-dimensional time-dependent Schr\"{o}dinger equation (TDSE). The TDSE is propagated using a symmetrized split-step operator method \cite{Steiger1982}. In neglecting atom-atom interactions we assume that the lattice is sparsely populated, e.g. though the use of a 3-D lattice where in the non-shaken dimensions the lattice is so deep that few atoms populate each site, but in the shaken dimension, the lattice is shallow enough that the atoms remain delocalized. The infinite lattice approximation used here is reasonable if the lattice beam has a Rayleigh range much larger than the lattice wavelength. This is true for lattice beams with near-infrared wavelengths and waists on the order of tens of microns.

Consider atoms trapped in a red-detuned optical lattice potential with a time varying phase $\phi(t)$. The potential is given by 
\begin{equation}
\label{eq.potential}
V(x,t) = -\frac{V_0}{2}\cos{\left[2k_{\mathrm{L}} x + \phi(t)\right]},
\end{equation}
where the lattice wavenumber is $k_{\mathrm{L}} = 2\pi/\lambda_\mathrm{L}$ for a lattice wavelength $\lambda_\mathrm{L}$. The lattice depth is $V_0$, given in units of the lattice recoil energy $E_\mathrm{R} = \hbar^2 k_\mathrm{L}^2/2m$ for an atom with mass $m$. In this work we take the lattice depth to be $V_0 = 10E_\mathrm{R}$. This depth is chosen because it is easily accessible experimentally and shallow enough that atoms remain delocalized in the lattice potential.

The main goal of SLI is to find an optimal shaking function $\phi(t)$ that transforms an initial state $\psi_\mathrm{0}$ to a desired final state $\psi_\mathrm{d}$. To accomplish this, we use a GA to optimize the shaking function to within a specified precision. In this way we construct a sequence of shaking functions that cause the atoms to undergo a conventional interferometer sequence.

\begin{figure}[]
\includegraphics[scale = 0.65]{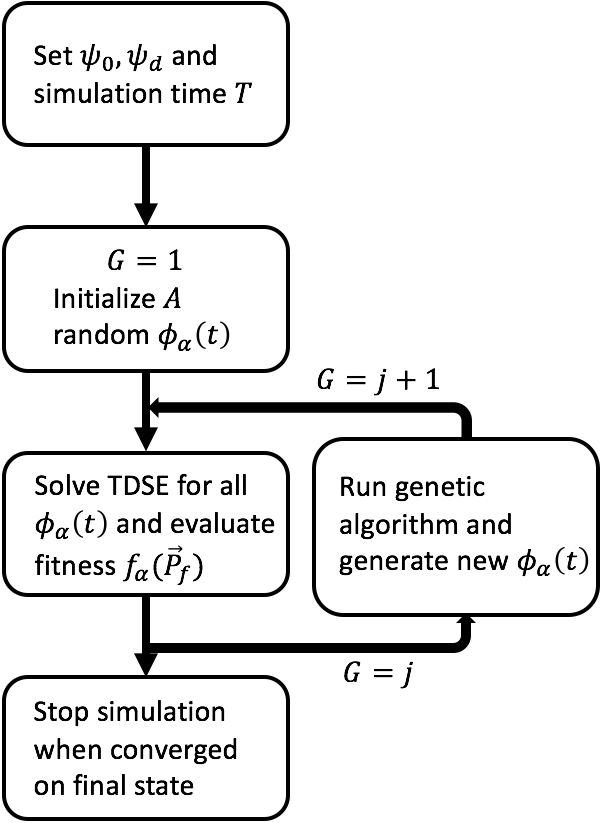}
\caption{\label{fig.Block} A block diagram illustrating the steps taken in the GA. Given the initial and desired states, the first generation $G = 1$ of $A$ individuals is randomly generated with unique shaking functions $\phi_\alpha(t)$. We then solve the TDSE for each individual in the generation. These results are then used to produce the next generation of individuals, denoted $G = 2$. After the $j^\mathrm{th}$ run of the simulation a generation $G = j + 1$ results from the mixing of the previous generation's individuals (see Fig. \ref{fig.GA_mut}). Once the convergence criterion is met, the GA stops.}
\end{figure}

The GA used in this work is based on the work in references \cite{Meystre2001, Rabitz1992} and relevant details are included here. A block diagram of the GA procedure is shown in Fig. \ref{fig.Block}. First, the algorithm produces an initial population of $A$ individuals. This is the first generation, denoted $G = 1$. Each individual $\alpha$ is produced by generating a random vector $\vec{\alpha}(\omega)$ of $l$ Fourier amplitudes at uniformly-spaced frequencies $\omega_i$ from DC up to a certain bandwidth set by the user. In this work, we set $A = 20$ and the Fourier amplitudes are randomly chosen from a normal distribution with standard deviation $\sigma = 100$. We pick $l \approx 100$ and limit the bandwidth to about $35$~kHz. The value of $\sigma$ is chosen to limit $\phi(t)$ to within approximately $\pm 2\pi$ radians and the bandwidth is chosen to limit the force on the atoms due to the shaking, as discussed further in Sec. \ref{sec.Sensitivity}. We choose an $l$ that gives good convergence of the GA within a reasonable amount of computation time.

Once the vector $\vec\alpha$ is chosen, its Fourier transform is taken to produce a time-varying function. To maintain smooth turn-on and turn-off, this function is multiplied by an envelope function $f_\mathrm{env}(t) = \sin^2{(\pi t/T)}$ that goes to zero at its endpoints. This results in a shaking function $\phi_\alpha(t)$ for each individual. The shaking time $T$ is a parameter chosen by the user and is set here to approximately $0.5$~ms. We choose the sine envelope shape over e.g, a Gaussian because it possesses a quicker ``turn-on'' time and in practice, the discontinuities at the endpoints do not cause undesired behavior in the shaking function.

Once the initial individuals are generated, the initial state $\psi_0$ is propagated forward in time by solving the TDSE with $\phi_\alpha(t)$, producing a final state $\psi_\alpha$. Because experimental measurements can provide momentum state populations via time-of-flight imaging, we represent a simulated result for the Fourier transform of the final wavefunction $\Psi_\alpha(k) = \mathcal{F}\{\psi_\alpha(x)\}$ as a vector $\vec{P}_\alpha(|\Psi_\alpha|^2)$. This vector has components $P_{\alpha,n}$ representing the relative atom population in each momentum state $2n\hbar k_\mathrm{L}$ for $n = [-\infty, \infty]$. A similar vector $\vec{P}_\mathrm{d}(|\Psi_\mathrm{d}|^2)$ is constructed for the desired final state. In this work $n$ is truncated at $n = [-5,5]$ since higher-order momentum states are negligibly populated. While it is possible to consider final states with non-integer momenta, these states remain unpopulated in our simulations and are thus not considered in the GA. This arises physically due to the fact that in momentum space the TDSE takes the form \cite{Meystre2001}
\begin{align}
\label{eq.TDSE_k}
i\hbar\frac{\partial\Psi(k,t)}{\partial t} &= \frac{\hbar^2 k^2}{2m}\Psi(k,t)- \frac{V_0}{4}[e^{i\phi(t)}\Psi(k-2k_\mathrm{L},t) \nonumber\\ & + e^{-i\phi(t)}\Psi(k+2k_\mathrm{L},t)].
\end{align}
Therefore, due to the presence of the applied potential, only transitions between momentum states separated by $2\hbar k_\mathrm{L}$ are allowed.

For a given shaking protocol $\phi_\alpha(t)$, the GA assesses the quality of the final momentum-space population via a fitness function $f(\vec{P}_\alpha)$ which quantifies the difference between the final and desired states. The fitness function will be discussed in detail later in this section. Once the fitness of the initial population is evaluated the individuals are ranked in terms of their fitness (because we are minimizing the fitness function, lower fitness values are better) and the genetic algorithm uses this ranking to produce the next generation of individuals. In general the procedure is identical from one generation $G = j$ to the next, $G = j+1$. The $A_\mathrm{live}$ best individuals are allowed to proceed unaltered to the next generation, a process known as ``elitism'' that ensures that the best (lowest) fitness value will never increase. The $A_\mathrm{die}$ worst individuals are deleted entirely. In this work we choose $A_\mathrm{live} = 2$ and $A_\mathrm{die} = 4$. The remaining $A - A_\mathrm{live}$ new individuals are generated by randomly picking ``parents'' from the surviving population for each new individual.

\begin{figure}[]
\includegraphics[scale = 0.93]{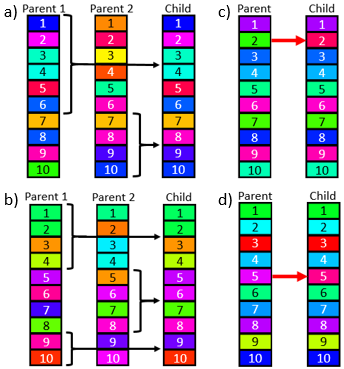}
\caption{\label{fig.GA_mut} (color online) Each iteration, the genetic algorithm mixes the best individuals from the previous generation to create ``children'' that populate the next generation. Each individual is a vector of values where red corresponds to a minimum value and purple corresponds to a maximum value. In each case the indices of crossover or mutation are chosen at random. The numbers label the index and their color is chosen for legibility. The methods presented here are adapted from reference \cite{Meystre2001}. a) One-point crossover at index $6$. b) Two-point crossover at indices $5$ and $9$. c) Random mutation of the value at index $2$. d) Change of the value at index $5$ through a small ``creep'' of the value.}
\end{figure}

The new generation is created from the previous generation using methods adapted from reference \cite{Meystre2001}. These methods are discussed here and illustrated in Fig. \ref{fig.GA_mut}. Given two parent vectors of Fourier amplitudes $\vec\alpha$ and $\vec\alpha'$, one-point crossover picks a random index $c$ and creates two new vectors $\vec\alpha_\mathrm{a}$ and $\vec\alpha_\mathrm{b}$ by swapping the values of the parent vectors starting at index $c$ such that $\vec\alpha_\mathrm{a} = \{\alpha_1,\alpha_2,...,\alpha_c,\alpha'_{c+1},...,\alpha'_l\}$ and $\vec\alpha_\mathrm{b} = \{\alpha'_1,\alpha'_2,...,\alpha'_c,\alpha_{c+1},...,\alpha_l\}$. Two-point crossover performs the same swap, but two indices $c_1$ and $c_2 > c_1$ are randomly chosen. This results in two children: $\vec\alpha_\mathrm{a} = \{\alpha_1,\alpha_2,...,\alpha_{c_1},\alpha'_{c_1+1},...,\alpha'_{c_2}, \alpha_{c_2+1}, ..., \alpha_l\}$ and $\vec\alpha_\mathrm{b} = \{\alpha'_1,\alpha'_2,...,\alpha'_{c_1},\alpha_{c_1+1},...,\alpha_{c_2}, \alpha'_{c_2+1}, ..., \alpha'_l\}$. Mutation takes a vector $\vec{\alpha}$ and produces a child $\vec{\alpha}_\mathrm{a}$. The child $\vec{\alpha}_\mathrm{a}$ is identical to $\vec{\alpha}$ except at a random index $c$ where $\alpha_{\mathrm{a},c} = \tilde{m}$ and $\tilde{m}$ is a random number such that $-M \leq \tilde{m} \leq M$ for some mutation limit $M$. Creep is similar to mutation, but $\alpha_{\mathrm{a},c} = \alpha_c + (0.5 - r)*C$, where $r$ is a random number between $0$ and $1$ and $C$ is defined as the ``creep'' rate. We set both $M$ and $C$ to be $1000$. These values are large enough to be effective at changing the population, but small enough to keep the final shaking function phase to within approximately $\pm 2\pi$ radians.

The idea behind the mixing in this and other GAs is to use the components of the $A - A_\mathrm{die}$ best individuals to produce a new generation with better fitness. When generation $j+1$ has been generated, the simulation runs again and evaluates the new fitness values. This procedure iterates until a preset number of iterations has been performed or the fitness reaches a suitable level.

The goal of the GA is to minimize the fitness function $f(\vec{P}_\alpha)$. The fitness function is constructed so that the final momentum state converges to the desired state, and the lattice ``learns'' to control the atoms. As an example we show the $f(\vec{P}_\alpha)$ for the split state in Eq. (\ref{eq.fit2hbark}). Splitting of the ground state wavefunction results in two traveling matter waves of the same amplitude with momenta of $\pm 2n\hbar k_\mathrm{L}$, analogous to an optical beamsplitter. For the simplest case modeled here the final state has momenta corresponding to $n = 1$. The fitness function used to evaluate the final split wavefunction is
\begin{align}
\label{eq.fit2hbark}
f(\vec{P}_\alpha) &= \vert\vec{P}_\mathrm{d} - \vec{P}_\alpha\vert + \sum\limits_{|n|\neq 1}\vert P_{\mathrm{d},n} - P_{\alpha,n}\vert\nonumber\\ &+
\bigg\vert\frac{P_{\alpha,1} - P_{\alpha,-1}}{P_{\alpha,1} + P_{\alpha,-1}}\bigg\vert.
\end{align}
The first term in Eq. (\ref{eq.fit2hbark}) quantifies the difference between the two momentum state populations for the final state and the desired state. The second term penalizes for atoms not in the desired $\pm 2 \hbar k_\mathrm{L}$ momentum states. The last term penalizes for asymmetry in the $\pm 2 \hbar k_\mathrm{L}$ momentum states. The penalizing terms are added to improve the rate of convergence to the desired state. Similar fitness functions are used for all shaking protocols developed in this work, and in general, any fitness function may be tailored to the individual requirements of each protocol in order to converge more quickly on a desired state.

\section{\label{sec.Mich}Development of a Michelson interferometer}

\begin{figure}[]
\includegraphics[scale = 0.73]{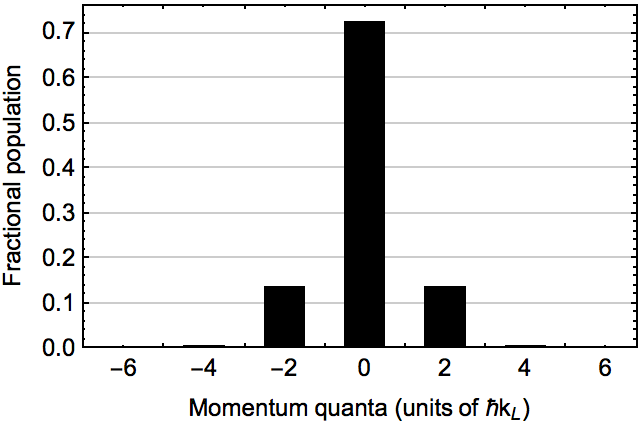}
\caption{\label{fig.init} The quantized momentum population of atoms in the ground Bloch state of an infinite lattice with $V_0 = 10 E_\mathrm{R}$. This ground state is the initial wavefunction of the atoms at the start of the interferometer sequence and the final state of the atoms after recombination with no signal applied.}
\end{figure}

\begin{figure}[]
\includegraphics[scale = 0.38]{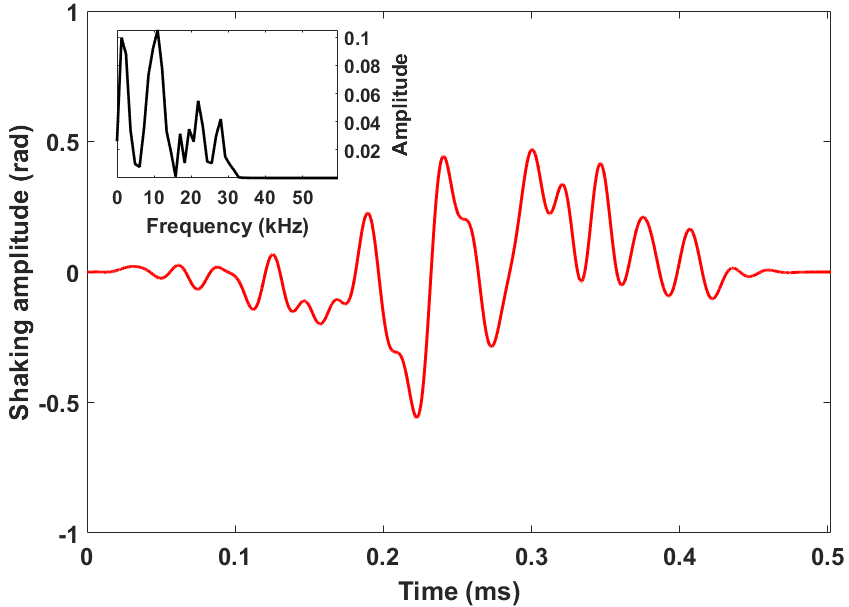}
\caption{\label{shaking_fcn} (color online) Shaking protocol for the optimal splitting shown in Table \ref{table:GA_results}. The envelope function ensures slow turn-on and turn-off at the endpoints. (inset) Fourier transform of the bandwidth-limited shaking function.}
\end{figure}

To implement the Michelson interferometer sequence we begin with atoms in the ground Bloch state of the lattice. The momentum population of the ground state is shown in Fig. \ref{fig.init} \cite{Phillips2002}. The GA then uses Eq. (\ref{eq.fit2hbark}) as the fitness function to find the optimal shaking protocol that splits the atom wavefunction. As stated in the last section the final split state is a pair of equal-amplitude waves moving with equal and opposite momenta. An optimized shaking protocol for splitting is shown in Fig. \ref{shaking_fcn}, showing the slow turn-on/off and the band-limited nature of the shaking function.

As the split states are not eigenstates of the lattice we find a second modulation protocol that maintains the purity of the momentum states as they propagate. Three additional shaking protocols are needed to reflect, reverse-propagate, and recombine the atoms. Once the five phase-modulation protocols are learned, interferometric measurements may be performed by repetition of their ordered sequence. The full Michelson interferometer sequence is illustrated in Fig. \ref{inter_full}.

During propagation and reflection it is important to control each momentum component separately because the initial and final momentum populations are the same. In these cases the TDSE solver is run twice. The first run begins with an initial state where all atoms are in the $-2n\hbar k_\mathrm{L}$ state and modulates the lattice with a function $\phi_\mathrm{pr}(t)$. The second run begins with atoms in the $2n\hbar k_\mathrm{L}$ state and applies the same phase modulation function $\phi_\mathrm{pr}(t)$. When propagating the atoms, the desired state is identical to the initial state for each momentum component. The GA then sums the fitness of both final states and optimizes $\phi_\mathrm{pr}(t)$. The optimal $\phi_\mathrm{pr}(t)$ will thus propagate the atoms without ``crosstalk'' between the two momentum states when applied to the linear combination of the two states. For reflection this simultaneous two-state optimization proceeds in the same way, but for an initial state with $\pm 2\hbar k_\mathrm{L}$, the final state has momentum $\mp 2\hbar k_\mathrm{L}$. After reflection the atoms are again propagated, this time with reversed momentum.

The final step of the interferometry sequence recombines the two split waves. The recombination scheme is as follows: the initial and desired states considered in splitting are swapped, such that the desired state is now the initial state and vice versa. The GA is run to find a modulation sequence that returns all of the atoms in the two split matter waves to the ground Bloch state. It is this final state of the atoms that changes when a signal is applied.

In this paper the variation $D_{j,k}$ between two states with momentum vectors $\vec{P}_j$ and $\vec{P}_k$ is defined as
\begin{equation}
\label{eq.Overlap}
D_{j,k} = (1 - \vec{P}_j\cdot \vec{P}_k)\times 100\%.
\end{equation}
By quantifying the difference $D_{f,a}$ between the final state after an acceleration signal is applied ($\vec{P}_\mathrm{a}$) and the final state after recombination ($\vec{P}_f$) one may quantify the degree of orthogonality between these two states. Equation (\ref{eq.Overlap}) provides a good measure of SLI robustness against noise for the purposes of this initial work. This is discussed further in Sec. \ref{sec.Robustness}.

For each shaking protocol, Table \ref{table:GA_results} shows the initial, desired, and final momentum states, as well as the difference between the optimized final state and the desired state as defined in Eq. (\ref{eq.Overlap}). The best result of 5 runs is shown, but in all cases, the variation is below $0.1\%$. Section \ref{sec.Sensitivity} will discuss the interferometer sensitivity and scaling with the interrogation time. Variation of the optimized results to variations in parameters will be discussed in the Sec. \ref{sec.Robustness}.

\begin{table*}[]
\caption{Genetic algorithm results, best of 5 optimization runs.} 
\renewcommand{\arraystretch}{1.2}
\centering 
\begin{tabular}{c c ccccc c c c c} 
\hline\hline 
Protocol & ~~State~~ & \multicolumn{5}{c}{Momentum population\footnotemark[1]} & ~~~$\%$ Diff.~~~ & ~~Bandwidth (kHz)\footnotemark[2]~~\\ 
\hline
 & & ~~$-4\hbar k_\mathrm{L}$~~ & ~~$-2\hbar k_\mathrm{L}$~~ & ~~~$0\hbar k_\mathrm{L}$~~~ & ~~~$2\hbar k_\mathrm{L}$~~~ & ~~~$4\hbar k_\mathrm{L}$~~~ & & & &\\
\hline 
Split & Init. & 0.0026 & 0.1345 & 0.7259 & 0.1345 & 0.0026 & & & & \\
 & Des. & 0 & 0.5 & 0 & 0.5 & 0 & & & & \\
 & Final & 0 & 0.4999 & 0.0001 & 0.4998 & 0 & $3.6\times 10^{-6}$ & 57.8\\ 
Prop. & Init. & 0 & 0.5 & 0 & 0.5 & 0 & & & & \\
 & Des. & 0 & 0.5 & 0 & 0.5 & 0 & & & & \\
 & Final & 0.0006 & 0.4992 & 0.0010 & 0.4980 & 0.0006 & $2.5\times 10^{-4}$ & 53.8 \\
Refl. & Init. & 0 & 0.5 & 0 & 0.5 & 0 & & & & \\
 & Des. & 0 & 0.5 & 0 & 0.5 & 0 & & & & \\
 & Final & 0.0012 & 0.4958 & 0.0037 & 0.4980 & 0.0008 & $1.8\times 10^{-3}$ & 55.8\\ 
Recomb. & Init. & 0 & 0.5 & 0 & 0.5 & 0 & & & & \\
 & Des. & 0.0026 & 0.1345 & 0.7259 & 0.1345 & 0.0026 & & & & \\
 & Final & 0.0026 & 0.1345 & 0.7258 & 0.1343 & 0.0026 & $5.4\times 10^{-6}$ & 55.8\\ 
\hline 
\footnotetext[1]{Normalized to 1. Momentum states higher than $|p| = 4\hbar k_\mathrm{L}$ are negligibly populated.}
\footnotetext[2]{Bandwidth is defined as the frequency where the amplitude drops below $30$~dB ($0.1\%$) of the maximum. The shaking time is $0.5$~ms.}
\end{tabular}
\label{table:GA_results}
\end{table*}

\section{\label{sec.Recip}Optimization of a reciprocal interferometer}

One of the advantages of SLI is the ability to control the transfer function of the interferometer. That is, we can change the shaking protocols to control the range and type of signals to which SLI is sensitive, such as an AC acceleration signal. Conversely, we can also design the interferometer to reject certain signals. In this section, we give examples of transfer function control with SLI.

To be sensitive to AC signals, the interferometer may be set up in a reciprocal configuration as shown in Fig. \ref{fig.inter_recip}. The difference between this and the interferometer configuration shown in Fig. \ref{inter_full} is that the atoms take a fully symmetric path. Because the space-time area of the interferometer is zero \cite{Kuhn2013}, the interferometer should be immune to DC accelerations but maximally sensitive to a sinusoidal acceleration
\begin{equation}
\label{eq.a}
\vec{a}(t) = a_x\sin{(\omega t)}\hat{x}
\end{equation}
at $\omega = 2\pi/T_\mathrm{I}$. Conversely, the non-reciprocal Michelson interferometer should be less sensitive to this acceleration but responsive to DC accelerations.

To show this the interferometer was simulated in the following ways: First, the standard Michelson interferometer was optimized so that the atom wavefunction was split, propagated for a time $2T_\mathrm{P}$, reflected, reverse propagated for a time $2T_\mathrm{P}$, then recombined into the ground state, as in Fig. \ref{inter_full}. The second simulation of the reciprocal interferometer (Fig. \ref{fig.inter_recip}) split the wavefunction, propagated for a time $T_\mathrm{P}$, reflected the atoms, reverse propagated them for a time $2T_\mathrm{P}$, reflected the atoms again, propagated them for a final time $T_\mathrm{P}$, then recombined them back into the ground state. In both cases the total propagation time was $4T_\mathrm{P}$. Once the simulations were completed, the shaking function $\phi_\mathrm{opt}(t)$ was used to simulate propagation of the TDSE with the potential
\begin{equation}
\label{eq.sine_pot}
V(x,t) = -\frac{V_0}{2}\cos{\{2k_\mathrm{L}x + \phi_\mathrm{opt}(t)\}} + m a_x x \sin{(\omega t)}.
\end{equation}

Note that any potential term linear in $x$ that is added to a lattice potential may be unitarily transformed into phase factor modifying the lattice potential (and vice versa) \cite{Zehnle2002}. Thus, from Eq. (\ref{eq.TDSE_k}), we expect that the momenta remain quantized even in the presence of an applied force, and this is verified by our simulations.

\begin{figure}[]
\includegraphics[scale = 0.485]{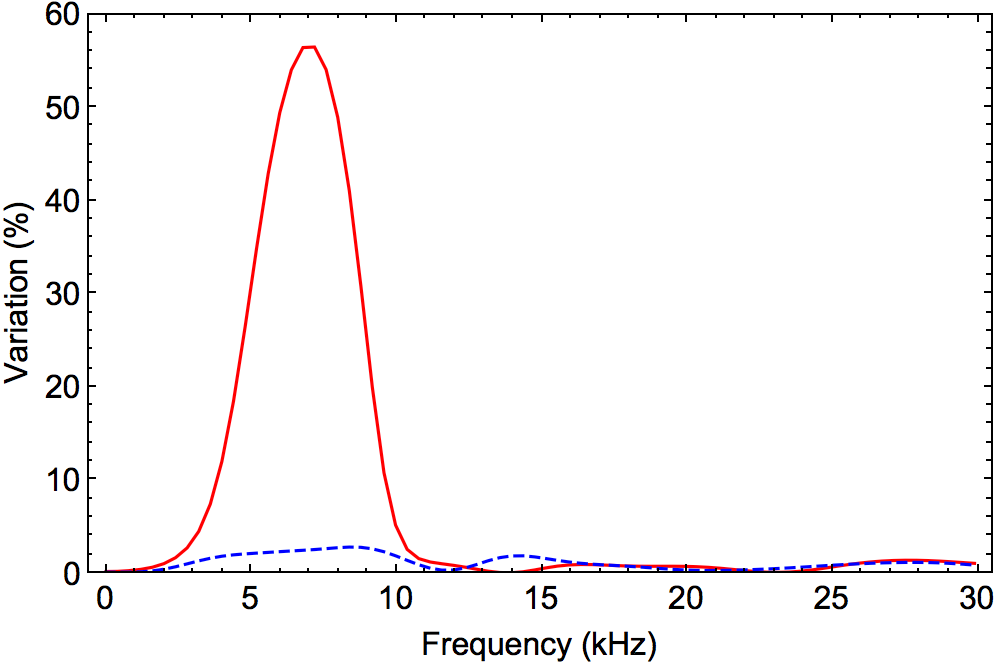}
\caption{\label{fig.recip} (color online) Response of the reciprocal (red, solid) and non-reciprocal (blue, dashed) interferometers to a sinusoidal signal as in Eq. (\ref{eq.sine_pot}). The response is given in terms of the variation $D_\mathrm{f,i}$ between the optimized final state of the interferometer and the final state after shaking with the applied signal as in Eq. (\ref{eq.Overlap}). The total interrogation time of each interferometer is $2.008$~ms and the amplitude of the applied acceleration signal is $a_0 = 0.115$~m/s$^2$. The reciprocal interferometer is $20$ times more sensitive than the non-reciprocal interferometer to a signal with $f\approx 7$~kHz, showing that SLI may be modified to tailor the interferometer response to an AC signal.}
\end{figure}

The phase of the acceleration during interrogation matters, and in this work, we analyze the effects of sinusoidal accelerations with no added phase. For the proof-of-principle simulations done here we set $T_\mathrm{P} = 0.502$~ms, $a_x = 0.115$~m/s$^2$, and scan the acceleration frequency from DC to $f = \omega/2\pi \approx 10/T_\mathrm{P}$ as shown in Fig. \ref{fig.recip}. We expect that the sensitivity of the reciprocal interferometer is maximal around $f = 1/4T_\mathrm{P} = 0.5$~kHz. Figure \ref{fig.recip} shows a maximum sensitivity for the reciprocal interferometer around $f_\mathrm{max} \approx 7$~kHz. Moreover the reciprocal interferometer shows a factor of $20$ enhancement in sensitivity over the non-reciprocal interferometer.

The discrepancy between simulation and theory in the value of $f_\mathrm{max}$ is due to the fact that the splitting, reflection, and recombination times are on the order of the propagation time, not negligibly small as assumed. Thus, the atoms interact with the shaken lattice for much longer than assumed, increasing the frequency at which the interferometer is maximally sensitive. Better agreement can be reached by increasing the propagation time relative to the splitting, reflection, and recombination times.

\begin{figure}[]
\includegraphics[scale = 0.375]{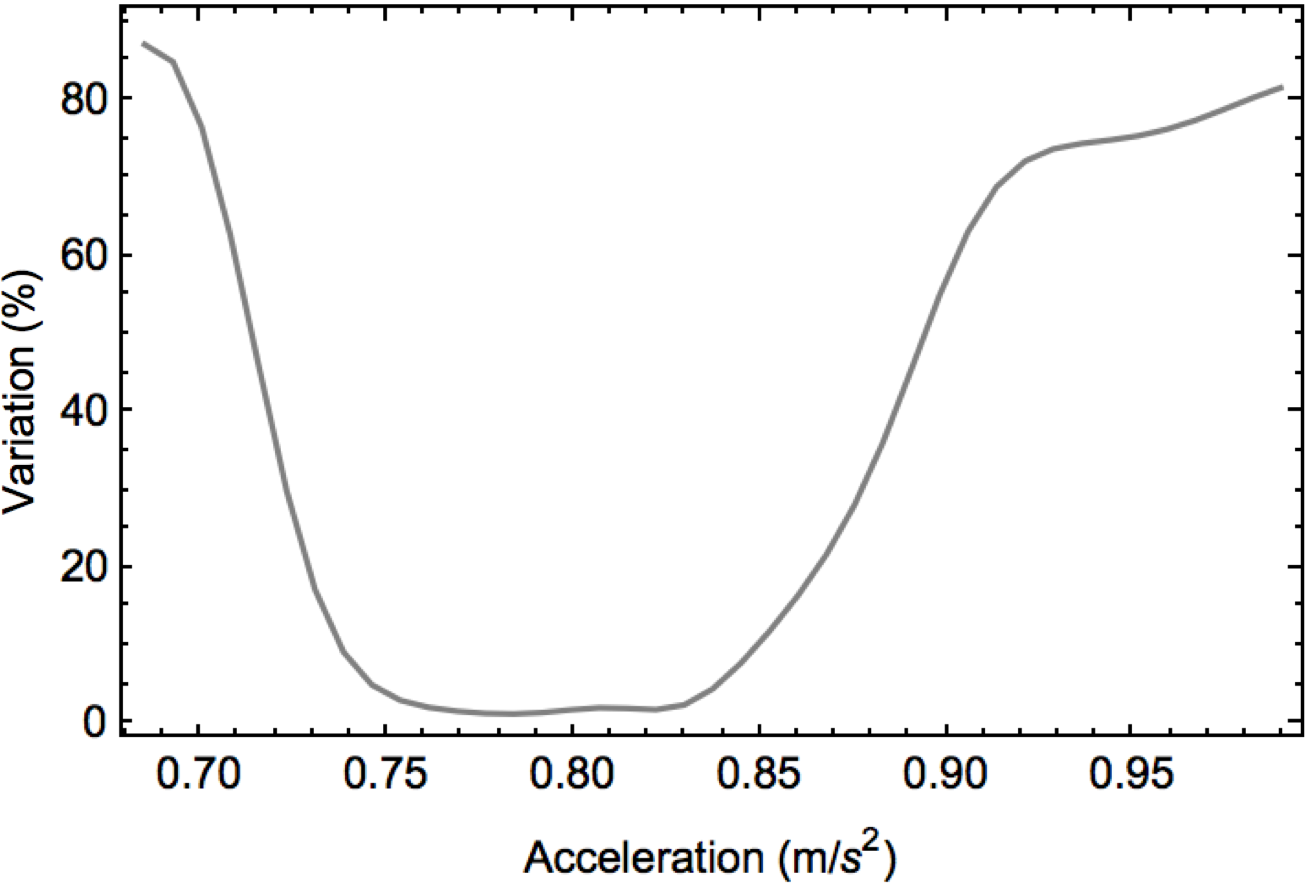}
\caption{\label{fig.bias} The response of an interferometer optimized in the presence of a DC bias acceleration $a_\mathrm{DC} = 0.76$ m/s$^2$. The interferometer response is clearly minimized in the vicinity of $a_\mathrm{DC}$ and increases away from this bias. This shows that SLI can be used to reject a DC bias of a given magnitude or measure perturbations around this bias.}
\end{figure}

The interferometer may be optimized with a DC bias so that it rejects a DC signal of a given magnitude. For the results shown in Fig. \ref{fig.bias} the standard non-reciprocal interferometer was optimized in the presence of a DC acceleration $a_\mathrm{DC} = 0.76$~m/s$^2$. Then the response of the interferometer was simulated for accelerations around $a_\mathrm{DC}$. Figure \ref{fig.bias} shows that the interferometer may be optimized and operated around a DC bias point. Thus, by combining this DC bias and AC sensitivity one can sense a time-varying acceleration of a certain frequency while rejecting a background DC acceleration. For example, a seismic signal could be detected while rejecting a DC signal due to gravity. An interferometer optimized in the presence of gravity at one spatial location could also be used to perform gravity gradiometry. A longer interrogation time will narrow the ``dip'' shown in Fig. \ref{fig.bias}, improving the interferometer response to signals around the bias. Currently, the only limits to total interrogation time are due to computation time and practical limits are due only to the experimental limitations discussed in Sec. \ref{sec.Sensitivity}. Future work will focus on increasing the interferometer sensitivity and optimization of the reciprocal interferometer for different values of the DC bias and AC signal acceleration.

\section{\label{sec.Sensitivity} Interferometer sensitivity and error}

In this section we calculate the interferometer sensitivity using the Fisher information metric \cite{Haine2016}. Then, possible sources of error in SLI are provided. These sources of error are considered in other conventional atom interferometry schemes, and potential solutions for the specific case of SLI are given. In many cases, if the lattice power does not drift and the optics are clean and stable, most errors can be corrected for in a closed-loop system.

We consider the Michelson and reciprocal interferometer schemes in this work as they are natural and familiar starting points. In general it is possible to optimize the interferometer to obtain higher sensitivities by changing the shaking protocol. Because the envelope function allows for smooth turn-on and turn-off of the shaking function, we can ``stitch'' together successive propagation steps to increase the interrogation time. Therefore the dynamic range of the interferometer can be controlled by changing the total propagation time. The GA is used to force the atom wavefunction to maintain its state over longer propagation times, correcting for any errors that arise as the propagation protocol is repeatedly applied.

In other atom interferometry schemes, a phase is measured between two atomic wavepackets. However, in SLI an applied acceleration will change the atoms' momenta as they interact with the shaken lattice. Therefore, because the atoms' momentum population changes under the influence of an applied signal, the definition of a phase difference between two arms becomes ambiguous because of the ambiguity in defining the arms of the interferometer.

It is due to this ambiguity that we use the classical Fisher information to quantify the interferometer sensitivity, given an optimized state $\vec{P}_\mathrm{f}$ and the final state under acceleration $\vec{P}_\mathrm{a}$. The classical Fisher information for a measured parameter $\theta$ given a probability distribution $f(x,\theta)$ can generally be written as
\begin{equation}
\label{eq.FI_class}
F_\mathrm{C}(\theta) = \int dx \bigg ( \frac{\partial}{\partial \theta}\ln{(f(x,\theta))}\bigg )^2 f(x,\theta).
\end{equation}
We may then use the Cramer-Rao bound to find the smallest resolvable change in the variable $\theta$. The Cramer-Rao bound is written
\begin{equation}
\label{eq.CR}
\delta\theta = \frac{1}{\sqrt{F_\mathrm{C}}}.
\end{equation}

In the simulation, we have access to the full atom wavefunction. Thus, we could calculate the quantum Fisher information \cite{Haine2016} for the acceleration parameter $a$. However, experimentally we only have access to amplitude information and lose information about relative phases between the different interfering momentum states. Therefore, in this work we only use the classical Fisher information. To simplify the problem, we will re-write Eq. (\ref{eq.FI_class}) using the momentum population vector $\vec{P}_\mathrm{a}$. In this case, we can write
\begin{equation}
\label{eq.FI_class_P}
F_{C,P}(a) = N_\mathrm{at}\sum_{n=-N}^N\,\frac{(\partial{P_{\mathrm{a},n}/\partial a)^2}}{P_{\mathrm{a},n}} = N_\mathrm{at}(\vec{A}\cdot\vec{B})
\end{equation}
where $\vec{A}$ has components $A_n = 1/P_{\mathrm{a}, n}$ and $\vec{B}$ has components $B_n = (\partial P_{\mathrm{a}, n}/\partial a)^2$ and $N_\mathrm{at}$ is the total number of atoms. The sum extends from $-N$ to $N$ where $N = 5$ is where we truncate the number of momentum states considered (see Sec. \ref{sec.Opt}) The factor $N_\mathrm{at}$ arises because if the atoms are non-interacting, each atom counted is a separate measurement of the probability distribution $\vec{P}_\mathrm{a}$. This will give us a factor of $\sqrt{N_\mathrm{at}}$ in the denominator of the expression for $\delta a$, as expected from conventional interferometry. The Fisher information $F_\mathrm{C,P}$ increases as $\partial P_{\mathrm{a}, n}/\partial a$ increases. This can be intuitively understood because operating in a regime with high $\partial P_{\mathrm{a}, n}/\partial a$ is analogous to operating on the edge of a fringe in conventional interferometry. From Eq. (\ref{eq.FI_class_P}), we can immediately write down the smallest resolvable acceleration $\delta a$ using Eq. (\ref{eq.CR})
\begin{equation}
\label{eq.CR_P}
\delta a = \frac{1}{\sqrt{N_\mathrm{at}}}\frac{1}{\sqrt{\vec{A}\cdot \vec{B}}}.
\end{equation}

To study how $\delta a$ changes with interrogtion time $T_\mathrm{I}$, we optimized the full interferometer for varying interrogation times from approximately $1$ to $20$~ms. Once the optimization was complete, we took the optimized shaking function $\phi_\mathrm{opt}(t)$ and added an acceleration signal, altering the potential from Eq. (\ref{eq.potential}) to
\begin{equation}
\label{eq.accel}
V(x,t) = -\frac{V_0}{2}\cos{\{2k_\mathrm{L}x + \phi_\mathrm{opt}(t)\}} + m a_x x.
\end{equation}
We then solved the TDSE with this potential and recorded the final momentum state vector $\vec{P}_\mathrm{a}$ for various values of the acceleration $a_x$. From this we use the Cramer-Rao bound to get $\delta a(T_\mathrm{I})$, where the derivatives were taken around $a = 0$.

\begin{figure}[]
\includegraphics[scale = 0.65]{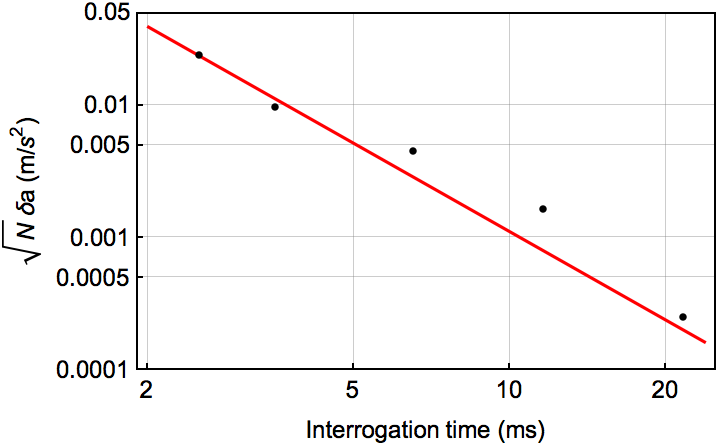}
\caption{\label{Fig.SLI_FI_init} The minimum detectable acceleration $\delta a$ scaled by $\sqrt{N_\mathrm{a}}$ plotted on a log scale versus the interrogation time $T_\mathrm{I}$. The black points are simulation results, and the red line is a fit of the form $C T_\mathrm{I}^{-n}$. Here, lower values of $\delta a$ correspond to a more sensitive interferometer.}
\end{figure}

The value of $\delta a$ should decrease as $T_\mathrm{I}$ increases such that $\delta a \propto T_\mathrm{I}^{-n}$ where $n$ is the interferometer scaling, since a lower value of $\delta a$ corresponds to an interferometer that is more sensitive to acceleration. To obtain $n$, we use the Levenberg-Marquardt algorithm to fit the resulting data to a curve of the form $f(T_\mathrm{I}) = C T_\mathrm{I}^{-n}$ for some constant $C$. The results are plotted in Figure \ref{Fig.SLI_FI_init}. Note that the total interrogation time also includes the nonzero splitting, reversal, and recombination time (about $1.5$~ms of total shaking time). The fit gives $n = 2.21 \pm 0.31$. This is consistent with the $n = 2$ scaling achieved in most other atom interferometers (e.g. light-pulse atom interferometers).

\begin{figure}[]
\includegraphics[scale = 0.65]{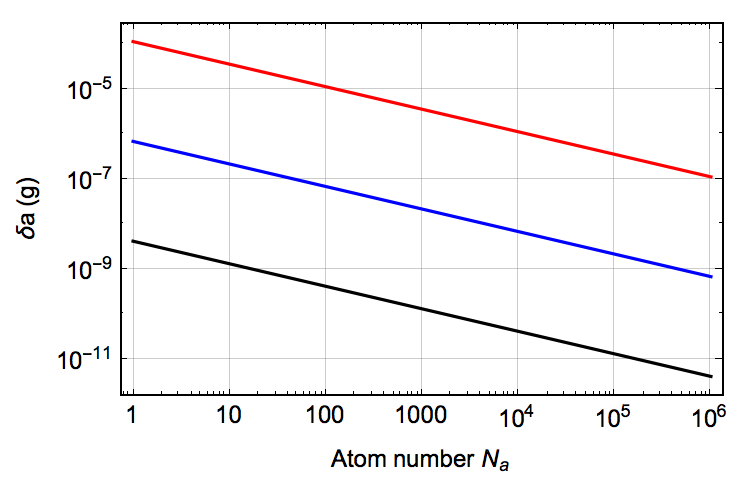}
\caption{\label{Fig.SLI_sens} The acceleration sensitivity $\delta a$ (relative to $g$) for varying atom numbers $N_\mathrm{a}$ at $T_\mathrm{I}=10$~ms (red), $100$~ms (blue), and $1$~s (black).}
\end{figure}

Using the numerical fit we can plot $\delta a$ versus the atom number for various interrogation times. This is shown in Fig. \ref{Fig.SLI_sens}. From this plot, we obtain $\delta a < 10^{-11}g$ for interrogation times of $1$~s and a million atoms.

It is possible that other shaken lattice interferometer configurations could scale with higher powers in $T_\mathrm{I}$. For example, one could optimize the interferometer to accelerate the atoms in the lattice as they propagate, much like the continuous-acceleration-Bloch interferometers that scale as $T_\mathrm{I}^3$ \cite{Robins2014}. One could also alter the fitness function so that the GA now minimizes the Cramer-Rao bound in Eq. (\ref{eq.CR_P}) and maximizes the interferometer sensitivity. This will be considered in a future work.

Trapped-atom interferometers generally suffer from the deleterious effects of phase diffusion due to atom-atom interactions \cite{Java1997}. In the shaken lattice interferometer, one may lower these interactions through sparse population of the lattice, e.g. via the use of the three-dimensional scheme discussed in Section \ref{sec.Opt}. That is, a deep two-dimensional lattice with low single-site occupation may be used. This results in an array of 1-D low atom number interferometers largely immune to the effects of phase diffusion. Shaking then takes place along the third dimension. If each of the 1-D interferometers is shaken in the same way, a collective measurement of their responses can be made. To counter the effects of shot noise, a few hundred of these 1-D ``tubes'' can be loaded with about 100 atoms each. Then total atom numbers can reach $10^6$, lowering shot noise to levels comparable with state-of-the-art light-pulse atom interferometers based on Bose-condensed atoms \cite{Kasevich2015}.

Using the results above for the sensitivity scaling with $T_\mathrm{I}$ we can plot the acceleration sensitivity of the interferometer versus the atom number for various interrogation times. This is shown in Fig. \ref{Fig.SLI_sens}. From this plot, we can expect a sensitivity better than $10^{-11}g$ for interrogation times of $1$~s and $10^6$ atoms.

In practice an optimized shaking function can be found computationally and adjustments may be made by running an experiment with the learning algorithm in a closed-loop scheme, which has been done before in cold and ultracold atom experiments \cite{Schumm2008, Hush2015, Zeilinger2016}. This may be used to optimize in the presence of inevitable systematics due to nonlinearities in shaking, laser wavefront errors, atom-atom interactions, and finite lattice effects. For example, a closed-loop system can correct deviations from optimal fitness due to the parasitic lattice reflections discussed in the previous section as long as the deleterious effects are constant from shot-to-shot. Uncertainties in lattice parameters such as the lattice depth or wavelength due to imperfect lattice alignment or the Gouy phase \cite{Bresson2012, Mueller2015} may also be corrected for in a closed-loop system.

As in light-pulse atom interferometry the effects of unwanted inertial signals, e.g. spurious rotations, can be subtracted out with the use of two interferometers operating in differential mode \cite{Rasel2009}. The common-mode signal can then be recovered by subtracting the two interferometer measurements.

The effects of decoherence from lattice shaking have been studied extensively elsewhere \cite{Arimondo2007, Tino2008, Ferrari2009, Arimondo2009}. These effects occur for a certain range of shaking amplitudes and frequencies, depending on the lattice depth and applied acceleration. If the desired dynamic range is known then undesired shaking frequencies and amplitudes can be filtered out accordingly in the learning algorithm.

A common limit to the lifetime of lattice experiments is defined by photon scattering. In other optical lattice experiments long atom lifetimes in the lattice are enabled by servo systems that lower laser noise \cite{Greiner2015}. If the lattice light is sufficiently far off of resonance the limit due to photon scattering is reduced (although more power is needed). Therefore, interrogation times on the order of tens of seconds are possible in shaken lattice systems. In the retro-reflecting lattice scheme, laser phase noise is irrelevant, but unwanted motion of the retro-reflecting mirror will cause unwanted shaking and give rise to spurious signals. Thus any noise in the retro mirror motion must be stabilized via a servo system.

Given the sources of error discussed in this section, we can also analyze the robustness of the shaken lattice interferometer. That is, we can study how the interferometer performance varies relative to variations in parameters such as lattice depth or wavelength. This is discussed in the next section.

\section{\label{sec.Robustness} Robustness of the interferometer}

In simulation the shaking function is optimized in an ideal situation where the lattice wavelength and depth are known exactly. However, in an experimental setting there is uncertainty in these values. Thus, the robustness of the shaking function to these errors is of interest. In this section analysis is done with respect to the optimal splitting function shown in Fig. \ref{shaking_fcn}, but the results presented here should hold for all stages of the interferometer due to the similarities in shaking time and bandwidth. As the interrogation time increases, however, we expect the stability requirements to become more stringent, and we present the results here only as a general example. In what follows, the variation $D_\mathrm{f,i}$ between the final state after perturbation and the optimized state is calculated using Eq. (\ref{eq.Overlap}).

\begin{figure}[]
\includegraphics[scale = 0.73]{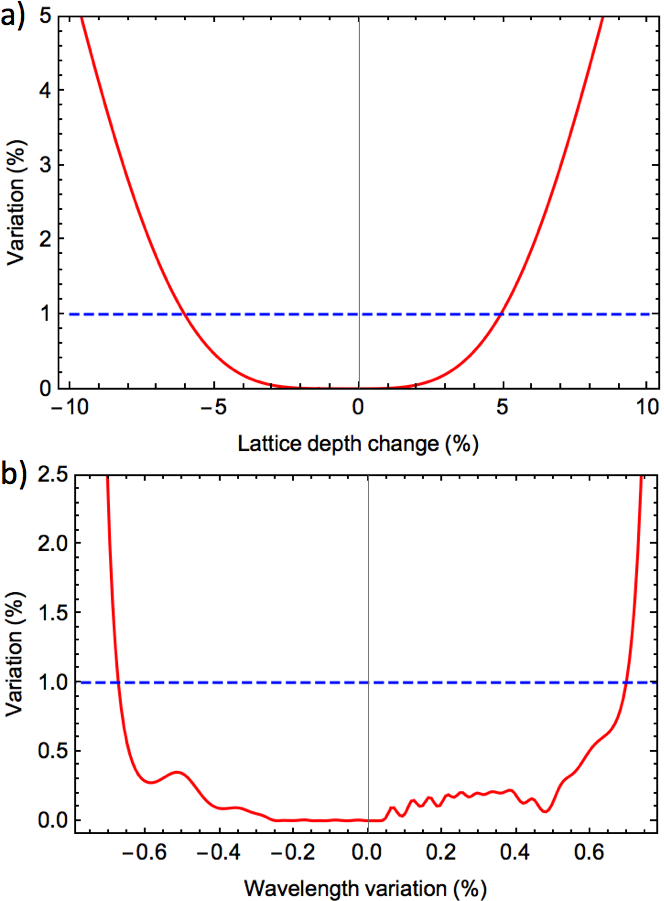}
\caption{\label{fig.Rob_Depth_Lambda} (color online) Percent variation of the optimized splitting shaking function shown in Fig. \ref{shaking_fcn} after variations of a) the simulated lattice depth and b) the simulated lattice wavelength. A variation of $1\%$ is marked in each plot with a blue dashed line.}
\end{figure}

The results of variation of lattice depth and wavelength are shown in Fig. \ref{fig.Rob_Depth_Lambda}. Changes in the lattice depth of $5\%$ and variations in the wavelength of $0.68\%$ maintain the variation between the perturbed final state and the optimized state to within $1\%$. For a mechanically stable system such lattice depth variations can be controlled by servoing the laser intensity, and as detailed in Sec. \ref{sec.Sensitivity} such a servo is desirable to limit heating due to laser intensity noise. Wavelength drift may be controlled by locking the laser, e.g. to an atomic hyperfine transition, providing frequency stability to better than $1$~MHz. Therefore the wavelength stability provided by laser locking is sufficient.

\begin{figure}[]
\includegraphics[scale = 0.145]{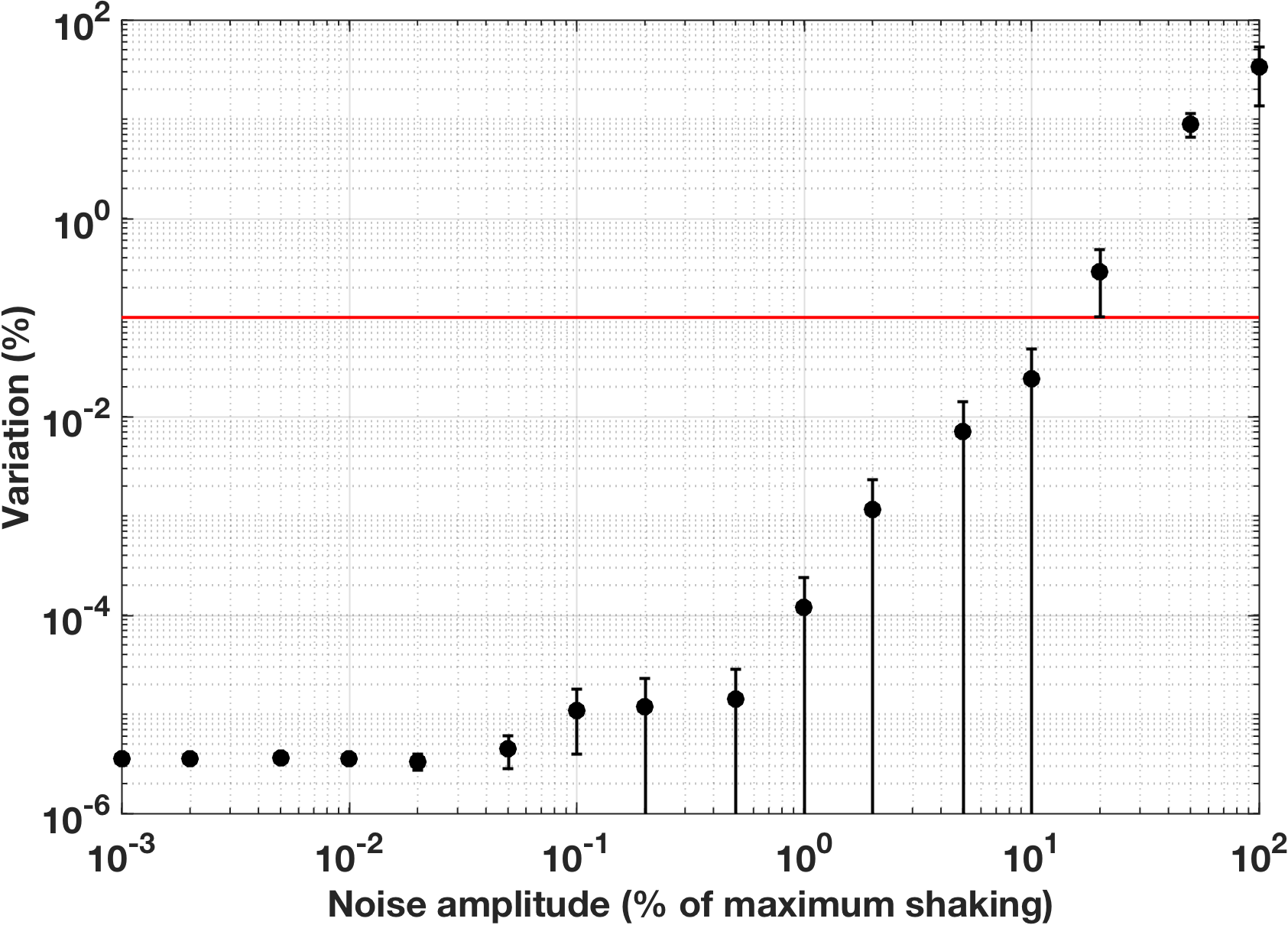}
\caption{\label{fig.Noise} (color online) Variation of the final state with varying noise amplitudes added to the optimized splitting shaking function. The results shown here are the average of $5$ runs with random white Gaussian noise added to the shaking function $\phi(t)$. Error bars give the standard deviation of the variations. Noise amplitude is given as a fraction of the maximum of the $\phi(t)$ shown in Fig. \ref{shaking_fcn}. The red line marks a variation of $1\%$.}
\end{figure}

Even if the lattice parameters are known exactly, in an experimental system the shaking will have some associated noise. For example, noise in the electronics used to shake (e.g. the driver of a piezoelectric mirror) will be added to the desired shaking function. Undesired shaking due to mechanical instability will also add noise to the optimal shaking function. To simulate this, Gaussian white noise of varying amplitudes was added to the optimized splitting function. The resulting variation in the perturbed final state relative to the optimized state is shown in Fig. \ref{fig.Noise}. At a threshold of about $10\%$, the variation rises rapidly as more noise is added to the shaking function. Thus, in a practical situation the noise in the shaking function should be kept below this threshold.

\begin{figure}[]
\includegraphics[scale = 0.68]{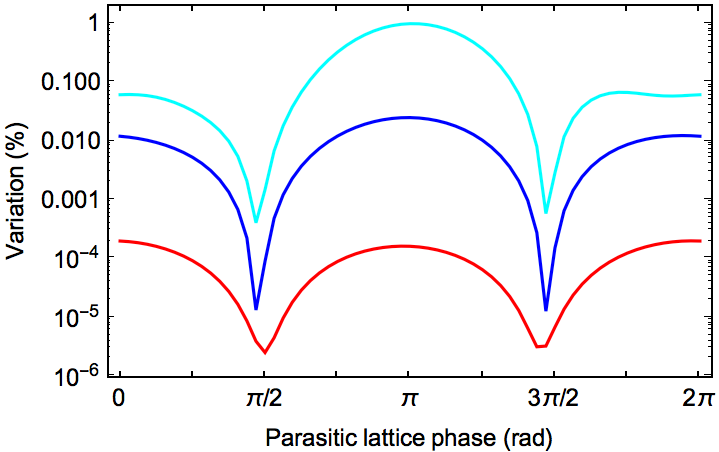}
\caption{\label{fig.ParasiticLatt} (color online) Variation of the final split state after the addition of spurious lattice potentials with varying phase due to unwanted reflections with reflection amplitudes of $\epsilon = 0.1\%$ (red), $1\%$ (blue), and $4\%$ (cyan).}
\end{figure}

Finally, we calculate the robustness of the shaking function to undesired stray reflections in the lattice system. In reference \cite{Bresson2012} the forces due to these reflections caused undesired phase shifts and imposed a limit on the contrast of the interferometer fringes. In our case such undesired reflections set up secondary lattice potentials that can cause the final state to deviate from the optimized result. We simulate this by shaking the lattice and adding a parasitic potential of the form
\begin{equation}
\label{eqn.parasitic_V}
V_\mathrm{P}(\epsilon, \delta) = -\frac{\epsilon V_0}{2}\cos{(2k_\mathrm{L}x + \delta)}.
\end{equation}
In Eq. (\ref{eqn.parasitic_V}), $\epsilon$ is the reflection amplitude and $\delta$ is the phase of the parasitic lattice relative to the initial unshaken main lattice potential. The value of $\delta$ is generally unknown in a real experiment. The simulation results are shown in Fig. \ref{fig.ParasiticLatt} for varying values of $\epsilon$ from $0.1\%$ (i.e. reflection from an anti-reflection coated window) to $4\%$ (reflection from uncoated glass). A single $4\%$ reflection can change the variation between the final state and optimal state by as much as $1\%$, and multiple such reflections should be managed carefully. Therefore, in an experimental realization, it is important to use AR coatings on the windows of the science chamber and align optics so that parasitic lattices do not interfere with the main lattice.
\protect\\
\section{\label{sec.Conc} Conclusion}

In conclusion, we introduce an atom interferometer based on a shaken lattice, where the shaking functions are found via a learning algorithm. This scheme provides the basis for a robust trapped-atom interferometer wherein the forces of interest are small perturbations on the shaking forces. The sensitivity of the interferometer may be increased by increasing the  interrogation time, which, if the atoms are trapped in the lattice potential, can be done practically without an increase in the size of the system. We show that SLI can be optimized to enhance signals of interest while simultaneously reducing the sensitivity to undesired signals. Several other aspects of shaken lattice interferometry will be studied in future work, such as increased scaling of the interferometer sensitivity relative to the shaking time and optimization to specific signals of interest, such as seismic accelerations. Finally, an experimental implementation will require a means of closed-loop learning to overcome inevitable experimental noise and systematic errors.

\begin{acknowledgments}
C.W, H.Y, and D.Z.A. would like to acknowledge funding from the Northrop Grumman Corporation and the NSF PFC under Grant No. 1125844.
\end{acknowledgments}

\bibliography{full_text_incl_figures.bbl}


\end{document}